\documentclass[12pt,pra,aps,amssymb,amsfonts,amsmath,tightenlines]{revtex4}
\usepackage{dsfont}
\usepackage{graphicx}
\usepackage{amssymb,amsfonts,amsthm}
\usepackage{color}
\usepackage[normalem]{ulem}

\newcommand{\half}{\mbox{$\textstyle \frac{1}{2}$}}

\newcommand{\rd}{\mbox{$\rm d$}}

\begin{document}

\title{Asset pricing with random information flow}

\author{Dorje~C.~Brody and Yan~Tai~Law}

\affiliation{Department of Mathematics, Imperial College
London, London SW7 2BZ, UK} 

\date{\today}

\begin{abstract}
In the information-based approach to asset pricing the market filtration 
is modelled explicitly as a superposition of signals concerning relevant 
market factors and independent noise. The rate at which the signal is 
revealed to the market then determines the overall magnitude of asset 
volatility. By letting this information flow rate random, we obtain an 
elementary stochastic volatility model within the information-based 
approach. Such an extension is economically justified on account of 
the fact that in real markets information flow rates are rarely measurable. 
Effects of having a random information flow rate is investigated in 
detail in the context of a simple model setup. Specifically, the price 
process of the asset is derived, and its characteristic behaviours are 
revealed via simulation studies. The price of a European-style option 
is worked out, showing that the model has a sufficient flexibility to fit 
volatility surface. As an extension of the random information flow model, 
price manipulation is considered. A simple model is used to show how 
the skewness of the manipulated and unmanipulated price processes 
take opposite signature. 
\end{abstract}


\maketitle

\section{Introduction}
\label{sec:1}

In the information-based asset pricing framework of Brody, Hughston and Macrina 
(hereafter the BHM framework) the starting point is the specification of a model for 
the market filtration, along with the cash flow of the asset \cite{bhm1,bhm2,M,RY}. 
The market filtration, 
more specifically, is generated by a \textit{market information process} that takes 
the form of a superposition of a `signal' component associated with the cash flow 
of the 
asset (or, more generally, market factors relevant to the actual cash flow) and an 
independent `noise' component that obscures the value of the cash flow. The 
simplest model for the information process within the BHM framework was 
introduced in the context of modelling credit-risky discount bond price process 
\cite{bhm1}. Specifically, we fix a probability space $(\Omega, {\mathcal F}_t, 
{\mathbb Q})$, where $\mathbb{Q}$ denotes the risk-neutral measure, and let 
$X_T$ denote the random variable representing the impending cash flow of 
a credit-risky bond, with maturity $T$. The process that generates market 
information is then defined by 
\begin{eqnarray}
\xi_t=\sigma X_T t + \beta_{tT} , \label{eq:1}
\end{eqnarray}
where $\{\beta_{tT}\}_{0\leq t \leq T}$ is a standard Brownian bridge on the interval 
$[0,T]$, independent of $X_T$, and $\sigma>0$ is constant. 

The rationale of the model (\ref{eq:1}) can be described briefly as follows. Before 
time $T$ market participants do not have direct access to the value of the  
cash flow. Market participants nevertheless have partial information (signal) 
concerning the value of $X_T$, which according to this simple model choice 
is revealed to the market at 
a constant rate $\sigma$. This `signal' is however obscured by an independent 
`noise', which is modelled here by a Brownian bridge $\{\beta_{tT}\}$. The market 
filtration $\{{\mathcal F}_t\}$ is thus identified to be that generated by the information 
process $\{\xi_{t}\}_{0\leq t \leq T}$. The price process of the asset that entails a 
single cash flow $X_T$ at time $T$, in this case might be viewed as a credit-risky 
discount bond, is thus obtained according to the prescription: 
\begin{eqnarray}
B_{tT} =P_{tT} {\mathbb E}^{\mathbb Q}[X_T | {\mathcal F}_t].
\end{eqnarray}
Here we let $\{P_{tT}\}$ denotes the discount function, which for simplicity is 
assumed deterministic. We thus see that, once the pricing measure ${\mathbb Q}$ 
is fixed, there are two inputs; cash flow $X_T$ and market filtration 
$\{{\mathcal F}_t\}$, the specification of which leads to an output that is the price 
process $\{B_{tT}\}$. 

The fact that it is market information that affects price dynamics has been 
emphasised by many authors (e.g., \cite{GS}), and has also been 
demonstrated against market data (e.g., \cite{ABDV,BBMP}). The aim of the 
BHM approach is therefore to bring the mathematical abstraction of financial 
modelling at the level of the specification of market filtration. In this way, 
price process can be derived as an emergent phenomenon, rather than 
postulated from the outset. 

One of the simplifying assumptions in the original BHM model (\ref{eq:1}) and 
the various generalisations of it that have appeared in the literature is that the 
information flow-rate parameter $\sigma$ is taken to be 
${\mathcal F}_0$-measurable. In a more realistic 
setup, however, market participants have little knowledge about the value of 
$\sigma$. In fact, in most cases the information flow rate is not measurable even 
after the value of $X_T$ is revealed. The main issue addressed in this paper is 
therefore 
to extend the original BHM model to allow for $\sigma$ to be a random variable. 
Such an extension constitutes the simplest stochastic volatility model within the 
BHM framework, which is shown to give sufficient flexibility to calibrate volatility 
surfaces. 

The paper is organised as follows. In \S\ref{sec:2} we begin by discussing the 
interpretation of the information process (\ref{eq:1}) as representing the totality 
of information available to the market concerning the value of $X_T$. In 
\S\ref{sec:2.5} we work out a statistical measure of sensitivity of the BHM model 
against the choice of the parameter $\sigma$, thus indicating the region in the 
parameter space that is susceptive to the misspecification of $\sigma$. 
In \S\ref{sec:3} we show that there is a kind of complementarity relation that holds 
between the information flow rate $\sigma$ and the cash flow $X_T$, and that 
they cannot both take arbitrary values owing to the measurability condition. In 
\S\ref{sec:4} we derive the expression for the asset price process in the random 
$\sigma$ environment. The associated price dynamics is worked out in 
\S\ref{sec:5}. In \S\ref{sec:6} and \S\ref{sec:7} we carry out a detailed numerical 
analysis to reveal a range of subtle details about the price dynamics. The 
pricing of an option is worked out in 
\S\ref{sec:8}. We conclude in \S\ref{sec:9} with a sketch of an idea on how 
we can model price manipulation in the information-based framework.

\section{Meaning of the information process}
\label{sec:2}

Before we proceed to investigate the properties of the model with a random 
$\sigma$ in detail, let us first comment on the meaning of the information 
process (\ref{eq:1}). In general, in a financial market, even if we make the 
simplifying assumption that 
the only relevant market factor is the cash flow $X_T$, there are plenty of 
information sources available for $X_T$, each being obscured by noise. We 
can therefore represent each of the information source in the 
signal-plus-noise form (\ref{eq:1}) and write 
\begin{eqnarray}
\left\{ \begin{array}{l} 
\xi^1_t = \sigma_1 X_T t + \beta^1_{tT} \\ 
\quad  \,\,\, \vdots  \\ 
\xi^n_t = \sigma_n X_T t + \beta^n_{tT} 
\end{array} \right. \label{eq:3}
\end{eqnarray}
for the family of information processes available in the market concerning the 
impending cash flow $X_T$. The various noise processes $\{\beta_{tT}^i\}_{
i=1,\ldots,n}$ in general may be mutually correlated (with correlation matrix 
$\rho$), but they are all independent of $X_T$. 

An important point to observe now is the fact that the aggregate information 
processes (\ref{eq:3}) is somewhat redundant; the information relevant to the 
cash flow $X_T$ contained in (\ref{eq:3}) can be represented in the form of 
a single information process (\ref{eq:1}), with the choice 
\begin{eqnarray}
\sigma^2 =\frac{\Sigma^n_i \sigma_i^2 \rho^{-1}_{ii} -2 
\Sigma_{i \neq j}\sigma_i \sigma_j \rho_{ij}^{-1}}{\det(\rho)} 
\label{eq:4}
\end{eqnarray}
for the \textit{effective information flow rate}, and
\begin{eqnarray}
\beta_{tT}=\frac{1}{\sigma}\left( \Sigma^N_{i,j} \sigma_i 
\rho^{-1}_{ij} \beta_{tT}^i \right)
\end{eqnarray}
for the \textit{effective noise}. 
Here $\rho^{-1}_{ij}$ denotes the $ij$ element of the 
inverse correlation matrix. 

Put the matter differently, the filtration generated jointly by the set of information 
processes (\ref{eq:3}) is equivalent to the filtration generated jointly by the single 
information process (\ref{eq:1}) and a family of noise processes given by 
combinations of $\{\beta_{tT}^i\}$. However, since noise terms are 
independent of $X_T$, they make no contribution to the pricing of the asset. 
We can therefore discard them altogether and represent the totality of 
`relevant' information in the form of a single information process (\ref{eq:1}). 

Remark: In the case of a pair of information processes on the same market 
factor $X_T$, the construction of an effective information process is used 
effectively in \cite{BDFH} to characterise the behaviour of an informed trader 
having access to additional noisy information.

\section{Sensitivity analysis}
\label{sec:2.5}

It is of interest to identify the sensitivity of the BHM model (\ref{eq:1}) to the 
specification of the information flow-rate parameter $\sigma$, given the fact 
that the value of $\sigma$ is usually unknown. Often one considers the 
option vega as a measure of parameter sensitivity, but here we are interested 
in a global measure of parameter sensitivity. The result will be useful in 
identifying the region in the parameter space for which a misspecification of 
the flow-rate parameter $\sigma$ yields significant errors in pricing a range 
of products, not just vanilla options. 

A universal measure of sensitivity in statistical analysis is given by the Fisher 
information \cite{Fisher}. To work out the Fisher information 
associated with the parameter $\sigma$ we proceed as follows. For simplicity, 
let as assume that the cash flow $X_T$ takes discrete values $\{x_i\}$ with 
\textit{a priori} probability $\{p_i\}$. To determine the information measure of 
Fisher we need the expression for the \textit{a posteriori} probability $\pi_{it}=
{\mathbb Q}(X_T=x_i|{\mathcal F}_t)$. It is shown in \cite{bhm1}, by making 
use of the Bayes formula, that this is given by 
\begin{eqnarray}
\pi_{it}=\frac{p_i \exp\left(\frac{T}{T-t}\left(\sigma x_i \xi_t - \frac{1}{2} 
\sigma^2 x_i^2 t\right) \right)}{\sum p_i \exp\left( \frac{T}{T-t}\left( 
\sigma x_i \xi_t - \frac{1}{2} \sigma^2 x_i^2 t\right) \right)}.
\end{eqnarray} 
We can therefore regard $\{\pi_{it}\}=\{\pi_{it}(\sigma)\}$ as a one-parameter 
family of probabilities. The Fisher information $g_t(\sigma)$ associated with 
the parameter $\sigma$ is then defined by the expression: 
\begin{eqnarray} 
g_t(\sigma) = \sum_i \frac{1}{\pi_{it}(\sigma)} 
\left( \frac{\partial \pi_{it}(\sigma)}{\partial \sigma} \right)^2.
\end{eqnarray} 

Remark: It is interesting that the Fisher information in the case of the BHM 
model (\ref{eq:1}) has the interpretation in terms of the conditional variance: 
\begin{eqnarray}
g_t(\sigma) = \sigma t \left({\textstyle\frac{T}{T-t}}\right)^2 
{\rm var} \left( X_T \beta_{tT} |\xi_t \right). \label{eq:x8}
\end{eqnarray}
To see this, let us define 
\begin{eqnarray}
p_{it}(\sigma) = p_i \exp\big({\textstyle\frac{T}{T-t}}\left(\sigma x_i \xi_t 
- \half \sigma^2 x_i^2 t\right) \big),
\end{eqnarray}  
so that $\pi_{it}=p_{it}/\sum_i p_{it}$. We then have 
\begin{eqnarray}
\frac{\partial p_{it}(\sigma)}{\partial \sigma} = \frac{T}{T-t} 
( x_i\xi_t - \sigma x_i^2 t) p_{it}(\sigma) ,
\end{eqnarray}  
from which it follows that 
\begin{eqnarray}
\frac{\partial \pi_{it}}{\partial \sigma} &=& \frac{1}{(\sum_i p_{it})^2} 
\left[ \frac{\partial p_{it}}{\partial \sigma} \sum_i p_{it} - p_{it} \sum_i 
\frac{\partial p_{it}}{\partial \sigma} \right] \nonumber \\ 
&=& \frac{T}{T-t}\, p_{it}\left( \frac{x_i \xi_t -\sigma x_i^2 t}{\sum_i 
p_{it}} - \frac{\sum_i p_{it}(x_i \xi_t -\sigma x_i^2 t)}{(\sum_i p_{it})^2} \right).
\end{eqnarray}  
Therefore, we obtain 
\begin{eqnarray}
g_t(\sigma) = \left(\frac{T}{T-t}\right)^2 \frac{\sum_i p_{it}\left( x_i \xi_t 
-\sigma x_i^2 t - \frac{\sum_i p_{it}(x_i \xi_t -\sigma x_i^2 t)}{\sum_i 
p_{it}}\right)^2}{\sum p_{it}} ,
\end{eqnarray}
and by substitution of (\ref{eq:1}) we deduce (\ref{eq:x8}). 

\begin{figure}[t]
\includegraphics[width=1\textwidth,clip]{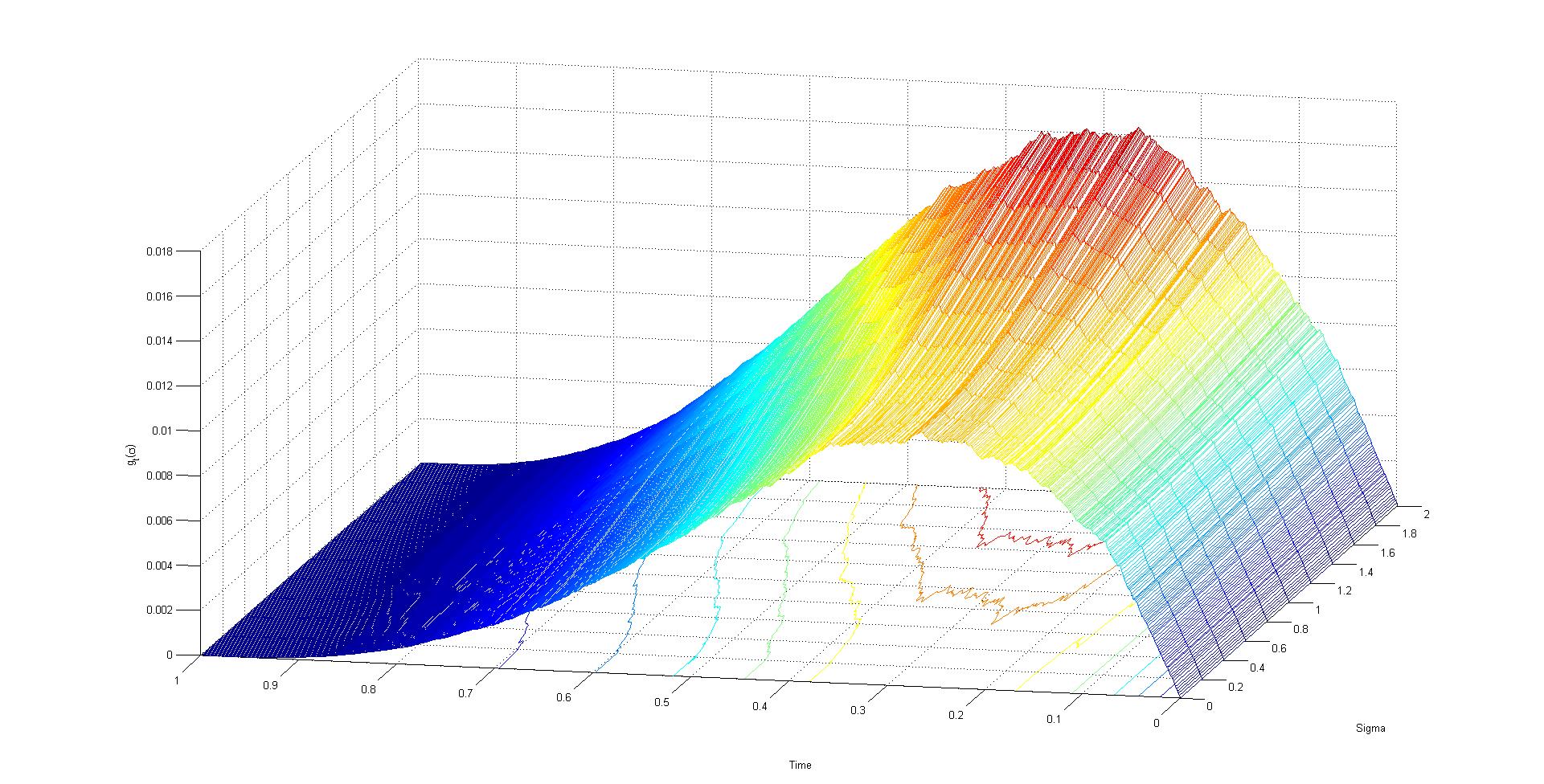}
\caption{The unconditional expectation ${\mathbb E}[g_t(\sigma)]$ of the 
Fisher information for different values of $t$. The numerical result is 
obtained by averaging over 1,000 sample paths. The parameters are chosen 
such that $\{x_i\}=\{0,0.5,1\}$, $\{p_i\}= \{0.1, 0.15,0.75\}$, and $T=1$.  
Where the value of ${\mathbb E}[g_t(\sigma)]$ is high, the model is 
on average sensitive to the choice of the specification of the parameter 
$\sigma$.}
\label{GSurface}
\end{figure}

The reason that $g_t(\sigma)$ measures the parameter sensitivity follows 
from the celebrated  Cram\'er-Rao inequality, which shows that the variance 
of the parameter estimate is bounded below by the inverse of the Fisher 
information. Additionally, as observed by Rao \cite{Rao}, the separation, i.e. 
the divergence measure associated with two models characterised by 
$\pi_{it}(\sigma)$ and $\pi_{it}(\sigma')$ is determined by the integral: 
\begin{eqnarray}
D_t(\sigma,\sigma') = \int^{\sigma'}_{\sigma}\!\! 
\sqrt{g_t(u)}\, \rd u.
\end{eqnarray}

In figure~\ref{GSurface} we show the numerical plot of the expectation 
${\mathbb E}[g_t(\sigma)]$ of the Fisher information for different values of 
$t$, averaged over 1,000 sample paths. The parameters are chosen such 
that $\{x_i\}=\{0,0.5,1\}$, $\{p_i\}= \{0.1, 0.15,0.75\}$, and $T=1$. We 
observe that the Fisher information is increasing in $\sigma$. It is evident 
from the plot that the model is quite sensitive to the choice of the information 
flow rate parameter $\sigma$ up until about two thirds of the way into the 
duration of the contract. The price of a generic derivative on a credit risky 
bond, with maturity not too close to zero but also shorter than two-third of the 
bond maturity, is thus likely to be sensitive to the specification of the 
parameter $\sigma$ in the most basic BHM model (\ref{eq:1}). This sensitivity 
can be made more robust by allowing $\sigma$ to be random.

\section{Quantisation of information flow rate}
\label{sec:3}

As indicated above, the complicated way in which the market information flow 
rate $\sigma$ depends 
on individual information flow rates $\{\sigma_i\}$, as represented in (\ref{eq:4}), 
suggests that it is not reasonable to assume that market participants have access 
to the value of $\sigma$. Indeed, in many realistic setup it is unlikely that market 
participants will ever learn the value of $\sigma$, even after the value of $X_T$ is 
revealed. More generally, the value of $\sigma$ can change over time in a random 
manner. Here we will be considering the simplest such situation in which 
$\sigma$ is given by a fixed random variable independent of $X_T$ and $\{
\beta_{tT}\}$. We observe, however, that whether $\sigma$ is a fixed random 
variable or a stochastic process, the ${\mathcal F}_T$-measurability of $X_T$ 
enforces a constraint on the allowable choice for $\sigma$. 

To see this, let us consider a simple example. Suppose that the random cash 
flow $X_T$ takes three possible values $\{0, 0.5, 1\}$ with \textit{a priori} 
probability $\{p_1,p_2,p_3\}$, and $\sigma$ takes two possible values 
$\{0.5,1\}$ with \textit{a priori} probability $\{q_1,q_2\}$. Then there are four 
possible realised values for the terminal information $\xi_T$: 
\begin{eqnarray}
\xi_T = \sigma T X_T = 
\begin{cases}
0 & \Rightarrow \, X_T=0, \quad \sigma=0.5~{\rm or}~1; \\
0.25T & \Rightarrow \, (X_T,\sigma) = (0.5,0.5); \\
0.5T & \Rightarrow \, (X_T,\sigma) = (0.5,1)~{\rm or}~(1,0.5); \\
T & \Rightarrow \, (X_T,\sigma)=(1,1). \\
\end{cases}
\label{eq:6}
\end{eqnarray}
In this case, if the realised value of $\xi_T$ happens to be $0$, $0.25T$, or 
$T$, then we can unambiguously determine the value of $X_T$ irrespective 
of what the value of $\sigma$ might have been, whereas if the outcome 
happens to be $0.5T$, then the value of $X_T$ could be $0.5$ or $1$. 

This example thus illustrates the fact that once the cash flow random variable 
is given, the information flow rate cannot take an arbitrary random variable. In 
particular, $\sigma$ can not be a continuous random variable. (More generally, 
if $\sigma=\sigma(t)$ is time dependent, then $\sigma(t)$ for $t<T$ can be 
arbitrary, but the constraint discussed here for the value of $\sigma(T)$ is 
still applicable.) Therefore, 
there is a kind of `quantisation' condition imposed on $\sigma$. For the same 
token, in the random-$\sigma$ environment, the cash flow $X_T$ cannot be 
a continuous variable---this does not pose real constraints because cash 
flows are in reality never continuous.

\section{Price process with a random information flow rate}
\label{sec:4}

Bearing in mind the quantisation condition imposed by the measurability of 
$X_T$, we proceed to consider the random-$\sigma$ extension of the BHM 
model. Specifically, we let $X_T$ take the values $\{x_i\}_{i=1,\ldots,n}$ with 
\textit{a priori} probabilities $\{p_i\}_{i=1,\ldots,n}$; and let $\sigma$ take the 
values $\{\sigma_k\}_{k=1,\ldots,m}$ with \textit{a priori} probabilities 
$\{q_k\}_{k=1,\ldots,m}$. These random variables are chosen such that 
degenerate situations like the example given in (\ref{eq:6}) are excluded, 
and hence $X_T$ is ensured to be ${\mathcal F}_T$-measurable. 

To determine the conditional expectation ${\mathbb E}[X_T|{\mathcal F}_t]$ 
we determine first the conditional probability $\pi_{it}={\mathbb Q}(X_T=x_i | 
{\mathcal F}_t)$. A calculation making use of the Bayes formula shows that 
this is given by 
\begin{eqnarray}
\pi_{it}= \frac{\sum^m_{k=1} p_i q_k \exp{\left[ \frac{T}{T-t} \left( x_i\sigma_k 
\xi_t - \frac{1}{2} x_i^2 \sigma_k^2 t\right)\right]}}{\sum^n_{s=1}\sum^m_{l=1} 
p_s q_l \exp{\left[ \frac{T}{T-t} \left( x_s\sigma_l \xi_t - \frac{1}{2} x_s^2 
\sigma_l^2 t\right)\right]}}. \label{eq:7}
\end{eqnarray}
It follows that the bond price is give by:
\begin{eqnarray}
B_{tT}= P_{tT}\frac{\sum^n_i\sum^m_{k=1} x_ip_i q_k \exp{\left[ \frac{T}{T-t} 
\left( x_i\sigma_k \xi_t - \frac{1}{2} x_i^2 \sigma_k^2 t\right)\right]}}
{\sum^n_{s=1}\sum^m_{l=1} p_s q_l \exp{\left[ \frac{T}{T-t} \left( x_s
\sigma_l \xi_t - \frac{1}{2} x_s^2 \sigma_l^2 t\right)\right]}}. \label{eq:8}
\end{eqnarray}

The derivation of (\ref{eq:7}) follows closely that of the original BHM model 
in \cite{bhm1}. First, we show that the information process (\ref{eq:1}) with a 
random $\sigma$ is a Markov process satisfying 
\begin{eqnarray}
{\mathbb Q}(\xi_t \leq x | \mathcal{F}_s )= {\mathbb Q} (\xi_t \leq x | \xi_s)
\end{eqnarray}
for all $x \in \mathbb{R}$ and all $s,t$ such that $0 \leq s \leq t \leq T$. To 
establish this, it suffices to show that 
\begin{eqnarray}
{\mathbb Q}(\xi_t \leq x | \xi_s, \xi_{s_1},..., \xi_{s_k} ) = {\mathbb Q} 
(\xi_t \leq x | \xi_s)
\end{eqnarray}
for any collection of times $t,s,s_1,...,s_k$ such that $T \geq t >s>s_1 >...>
s_k>0$. We recall first that from properties of a Brownian bridge process it 
follows that $(\beta_{sT}/s-\beta_{s_1T}/s_1)$ and $(\beta_{s_2T}/s_2-
\beta_{s_3T}/s_3)$ are independent. We now observe that 
\begin{eqnarray}
\mathbb{Q}( \xi_t \leq x \mid \xi_s, \xi_{s_1},\cdots,\xi_{s_k})&=& 
\mathbb{Q} \left( \xi_t \leq x \mid \xi_s, \frac{\xi_s}{s}- \frac{\xi_{s_1}}{s_1}, 
\cdots, \frac{\xi_{s_{k-1}}}{s_{k-1}}- \frac{\xi_{s_k}}{s_k} \right) \nonumber \\
&=& \mathbb{Q} \left( \xi_t \leq x \mid \xi_s, \frac{\beta_{sT}}{s}- 
\frac{\beta_{s_1T}}{s_1} , \cdots, \frac{\beta_{s_{k-1}T}}{s_{k-1}}- 
\frac{\beta_{s_kT}}{s_k} \right),
\end{eqnarray}
but since $\xi_s$ and $\xi_t$ are independent of the remaining variables 
$\beta_{sT}/s-\beta_{s_1T}/s_1$, $\cdots$, $\beta_{s_{k-1}T}/s_{k-1}- 
\beta_{s_kT}/s_k$, the desired Markov property follows.

From the Markovian property of $\{\xi_t\}$ the problem of determining the 
conditional probability process $\{\pi_{it}\}$ simplifies to calculating the 
conditional probability $\mathbb{Q}(X_T=x_i | \xi_t)$. Then from the Bayes 
formula we find 
\begin{eqnarray}
{\mathbb Q}(X_T=x_i | \xi_t) = \frac{{\mathbb Q}(X_T=x_i)
\rho(\xi_t | X_T=x_i)}{\sum_i {\mathbb Q}(X_T=x_i)
\rho(\xi_t | X_T=x_i)} , 
\end{eqnarray}
where $\rho(\xi|X_T=x_i)$ is the conditional density for $\xi_t$. But from 
\begin{eqnarray}
\rho(\xi_t | X_T=x_i) = \sum_{k} \rho(\xi_t | X_T=x_i, \sigma
=\sigma_k) {\mathbb Q}(\sigma=\sigma_k)
\end{eqnarray}
we deduce that 
\begin{eqnarray}
\pi_{it} = \frac{\sum_k p_i q_k \rho
(\xi_t | X_T=x_i, \sigma= \sigma_k)}{ \sum_i \sum_k p_i q_k 
\rho(\xi_t | X_T=x_i, \sigma= \sigma_k)} . \label{eq:14} 
\end{eqnarray}
Conditional on $X_T=x_i$ and $\sigma = \sigma_k$ the random variable 
$\xi_t = \sigma X_T t + \beta_{tT}$ has the probability law of a drifted 
Brownian bridge: 
\begin{eqnarray}
\rho(\xi_t | X_T=x_i, \sigma= \sigma_k)= 
\frac{1}{\sqrt{2\pi t(T-t)/T}}\exp\left( -\frac{1}{2}
\frac{(\xi_t-\sigma_kx_it)^2}{t(T-t)/T}\right). \label{eq:15} 
\end{eqnarray}
Substituting (\ref{eq:15}) in (\ref{eq:14}) we deduce the desired expression 
(\ref{eq:7}).

\section{Price dynamics}
\label{sec:5}

With the expression (\ref{eq:8}) for the price process at hand  we are able 
to investigate its dynamics. To proceed let us write $p_{ikt}=p_{ik}(t,\xi_t)$, 
where
\begin{eqnarray}
p_{ik}(t,\xi) = p_i q_k \exp{\left[ \frac{T}{T-t} \left( 
x_i\sigma_k \xi - \half x_i^2 \sigma_k^2 t\right)\right]}.
\end{eqnarray}
Then an application of Ito's lemma gives 
\begin{eqnarray}
\frac{\rd p_{ikt}}{p_{ikt}} = \frac{\sigma_k T}{T-t}\, x_i 
\left(\textrm{d}\xi_t+ \frac{\xi_t}{T-t} \textrm{d}t \right) . 
\label{eq:17} 
\end{eqnarray}
It follows that the process $\{p_{it}\}$ defined by 
\begin{eqnarray}
p_{it} = \sum_k p_{ikt} \label{eq:18} 
\end{eqnarray}
fulfils the stochastic equation 
\begin{eqnarray}
\rd p_{it} = \frac{T}{T-t}\, x_i \left(\sum_k \sigma_k p_{ikt}\right) 
\left(\textrm{d}\xi_t+ \frac{\xi_t}{T-t} \textrm{d}t \right) . \label{eq:19} 
\end{eqnarray}

Since the \textit{a posteriori} density is given by 
\begin{eqnarray}
\pi_{it} = \frac{p_{it}}{\sum_i p_{it}}, \label{eq:20} 
\end{eqnarray}
another application of Ito's rule gives 
\begin{eqnarray}
\rd \pi_{it} = \frac{T}{T-t}\, \left(x_i \frac{\sum_k \sigma_k p_{ikt}}
{\sum_ip_{it}} - \pi_{it} {\mathbb E}[\sigma X_T |\xi_t] \right)  \rd W_{t},
\end{eqnarray}
where  
\begin{eqnarray}
\rd W_t = \rd \xi_t+ \frac{1}{T-t}\, (\xi_t - T {\mathbb E}[\sigma X_T |\xi_t]) \rd t . 
\label{eq:22} 
\end{eqnarray}
Putting these together, we find that the dynamical equation satisfied by 
the defaultable discount bond price is given by 
\begin{eqnarray}
\rd B_{tT} = r_t B_{tT} \rd t + \Sigma_{tT} \rd W_t, 
\end{eqnarray}
where the volatility process is determined by the conditional covariance 
process of $X_T$ and $\sigma X_T$: 
\begin{eqnarray}
\Sigma_{tT} = P_{tT} \frac{T}{T-t}\, {\rm cov}(X_T,\sigma X_T|\xi_t) . 
\end{eqnarray}
In particular, if $\sigma$ is constant, the covariance reduces to the variance 
of $X_T$, and we recover the original BHM model. We remark that although 
$\sigma$ and $X_T$ are \textit{ a priori} independent, they are not 
conditionally independent, and hence the covariance term does not reduce 
to a simpler expression ${\mathbb E}[\sigma|\xi_t]\, {\rm var}(X_T|\xi_t)$. 

Next, we establish that the process $\{W_t\}$ defined in (\ref{eq:22}) is 
the innovations representation associated with the filtering problem 
corresponding to the information process $\{\xi_t\}$. We shall follow 
closely the argument of \cite{bhm1} but extended to a random $\sigma$. 
Since $(\rd W_t)^2 
= \rd t$, to show that $\{W_t\}$ is a Brownian motion it suffices to 
establish that it is a martingale. For $t \leq u$ we have
\begin{eqnarray}
{\mathbb E}[W_u |\mathcal{F}_t] &=& W_t + {\mathbb E}[(\xi_u-\xi_t)|\xi_t] 
+ {\mathbb E}\left[\left. \int^u_t \frac{\xi_s}{T-s}\rd s \right| \xi_t \right] - T 
{\mathbb E}\left[ \left. \int^u_t \frac{{\mathbb E}[\sigma X_T |\xi_s]}{T-s} \rd s 
\right| \xi_t \right]  \nonumber\\
&=& W_t + {\mathbb E} \left[ \sigma X_T u + \beta_{uT}|\xi_t \right] - 
{\mathbb E} \left[ \sigma X_Tt + \beta_{tT}|\xi_t \right] + {\mathbb E} 
[\sigma X_T | \xi_t] \int^u_t \frac{s}{T-s}\rd s \nonumber \\ && + 
{\mathbb E}\left[\left. \int^u_t \frac{\beta_{sT}}{T-s}\rd s \right| \xi_t \right] - 
{\mathbb E}[\sigma X_T |\xi_t] \int^u_t \frac{T}{T-s}\, \rd s. \label{eq:25} 
\end{eqnarray}
Evidently, the coefficients of ${\mathbb E}[\sigma X_T |\xi_t]$ cancel, and 
we are left with 
\begin{eqnarray} 
\mathbb{E}[W_u|{\mathcal F}_t] = W_t + {\mathbb E}[ \beta_{uT}|\xi_t] - 
{\mathbb E}[\beta_{tT}| \xi_t] + \int^u_t \frac{1}{T-s}\, {\mathbb E} 
[\beta_{sT}|\xi_t] \rd s, \label{WMart2}
\end{eqnarray}
but because 
\begin{eqnarray}
{\mathbb E}[\beta_{uT}| \xi_t] =  \frac{T-u}{T-t}\, {\mathbb E}
[\beta_{tT} |\xi_{t}], 
\end{eqnarray}
we deduce the martingale condition
\begin{eqnarray}
{\mathbb E}[W_u | {\mathcal F}_t] = W_t.
\end{eqnarray}
It follows that the process $\{W_t\}$ defined via (\ref{eq:22}) is indeed a 
${\mathbb Q}$-Brownian motion with respect to the filtration generated 
by the information process $\{\xi_t\}$.

\section{Analysis of the sample-path behaviour}
\label{sec:6}

In this section we analyse the sample-path behaviour of a defaultable digital 
bond price under our uncertain information model through simulation studies. 
This provides us with a better intuitive understanding of the characteristics of 
the model under study. 

\begin{figure}[h!]
\centering
\includegraphics[width=1\textwidth,clip]{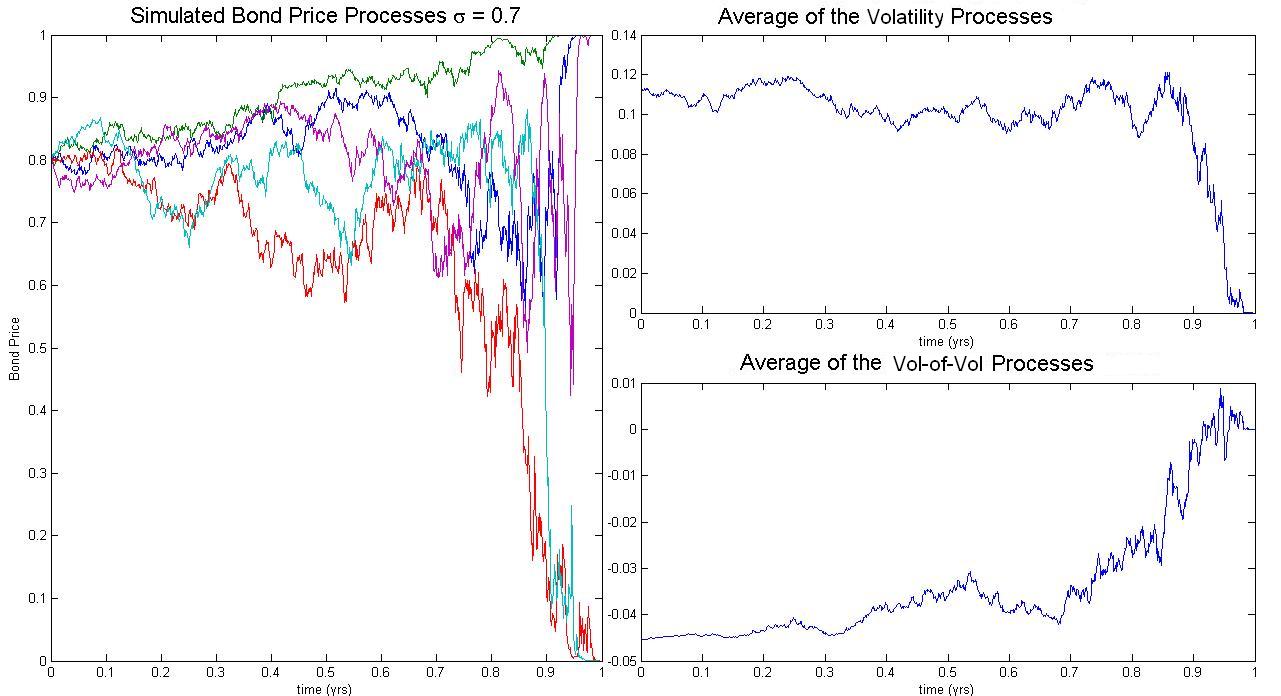}
\caption{Simulated Bond price (left), volatility (top right) and Vol-of-Vol 
(bottom right) processes in uncertain information model with 
$\sigma=\{0.6, 0.8\}$, $q_k=\{0.5,0.5\}$, $X_T=\{0,1\}$, $p_i=\{0.2,0.8\}$, 
$r=0\%$ and $T=1$.}
\label{Rand0608}
\end{figure}

\begin{figure}[h!]
\centering
\includegraphics[width=1\textwidth,clip]{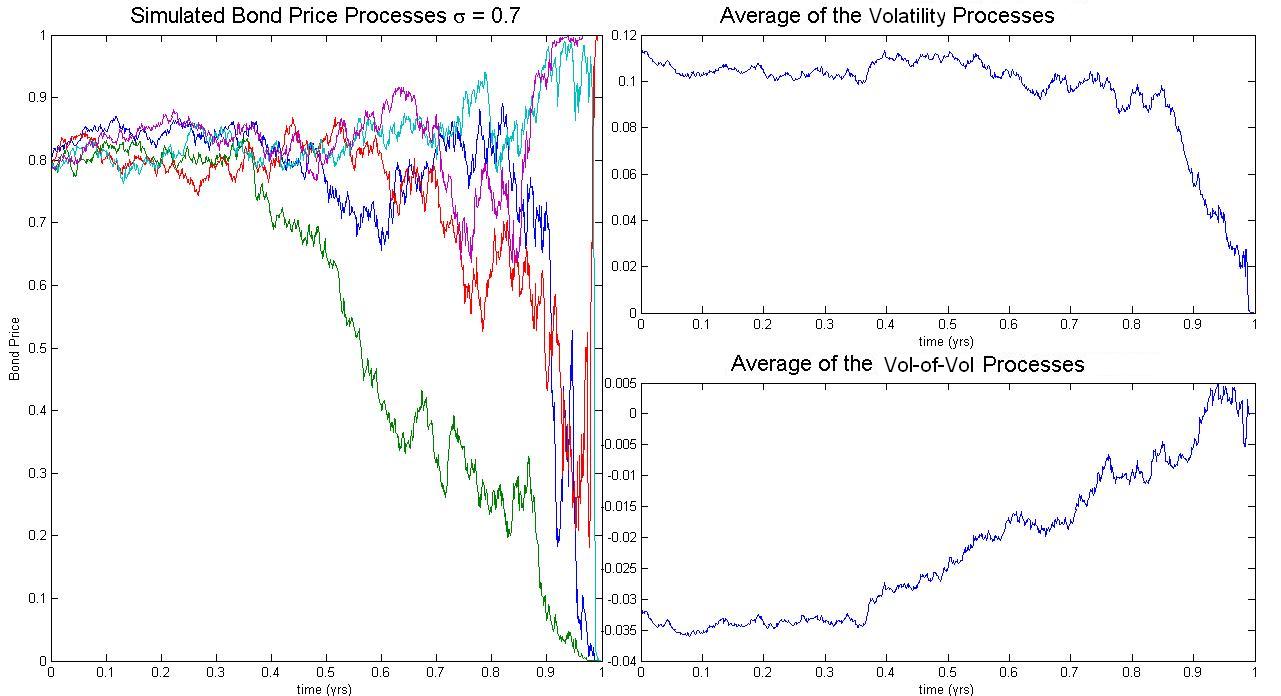}
\caption{Simulated Bond price (left), volatility (top right) and Vol-of-Vol 
(bottom right) processes in uncertain information model with 
$\sigma=\{0.4, 1\}$, $q_k=\{0.5,0.5\}$, $X_T=\{0,1\}$, $p_i=\{0.2,0.8\}$, 
$r=0\%$ and $T=1$.}
\label{Rand0410}
\end{figure}

In figure \ref{Rand0608} and \ref{Rand0410} we have shown simulations of 
sample paths of the bond price processes, the corresponding averaged 
(over five sample paths) volatility process, and the averaged (again, over five 
sample paths) vol-of-vol process. In these plots, we have set $r=0\%$ for 
simplicity; maturity of the digital bond is $T=1$. The information flow rate in 
figure~\ref{Rand0608} is chosen to be a binary random variable taking values 
$\{0.6,0.8\}$ with an equal \textit{a priori} probability; whereas in 
figure~\ref{Rand0410}, $\sigma$ takes the values $\{0.4,1.0\}$ also with an 
equal \textit{a priori} probability. Hence in both cases we have ${\mathbb E}
[\sigma]=0.7$, but they have different standard deviations. The 
simulation study shows that as we increase the variance of $\sigma$ the 
variance of the price paths increases. This behaviour is intuitively expected 
although not immediately apparent from the formula for the bond volatility. 

One other interesting observation to be drawn from the simulation studies, as 
compared to the original BHM model, is that the degree of variation of the 
sample paths, or simply the volatility, is smaller than those observed in the 
BHM with the same $\sigma$ value. For example, although one of the 
sample paths in figure \ref{Rand0410} has the value $\sigma=1$, which in 
the BHM model would have caused the paths to reach its terminal value at 
about four-fifth of the way, such a large variation is not present in our 
uncertain information model. This is because the conditional expectation of 
$X_T$ involves products of the \textit{a priori} probability $\{q_k\}$, and this 
`dampens' the overall variability. If we set one of the $q_k$'s equal to zero, 
which reduces the model to the original BHM, then the damping effect 
disappears. This behaviour is plausible because market participants are 
uncertain about the true information flow rate; their knowledge of the 
information is made additionally fuzzy by the uncertainty in $\sigma$, and 
hence it generally takes a longer time to discover the true terminal cash flow. 
This uncertainty will, consequently, reduce the overall volatility. The reduction 
in volatility becomes apparent when we look at the average of the processes. 

The volatility of the processes in some sense reflects the `learning rate' of the 
terminal value of the bond price (or, equivalently, the value of $X_T$). This 
learning rate is somewhat slowed down in the uncertain information model, 
hence resulting in the dampening of the volatility. If the standard deviation of 
$\sigma$ is very small, then the volatility process almost matches that of the 
BHM model. However, as the standard deviation of $\sigma$ is made wider 
so that the true information becomes less clear, this leads to the reduction in 
the price volatility. Nevertheless, as time passes, the value of $X_T$ must 
eventually be revealed, and this leads to a subtle behaviour in the averaged 
volatility. 

\begin{figure}[htp]
\centering
\includegraphics[width=0.9\textwidth,clip]{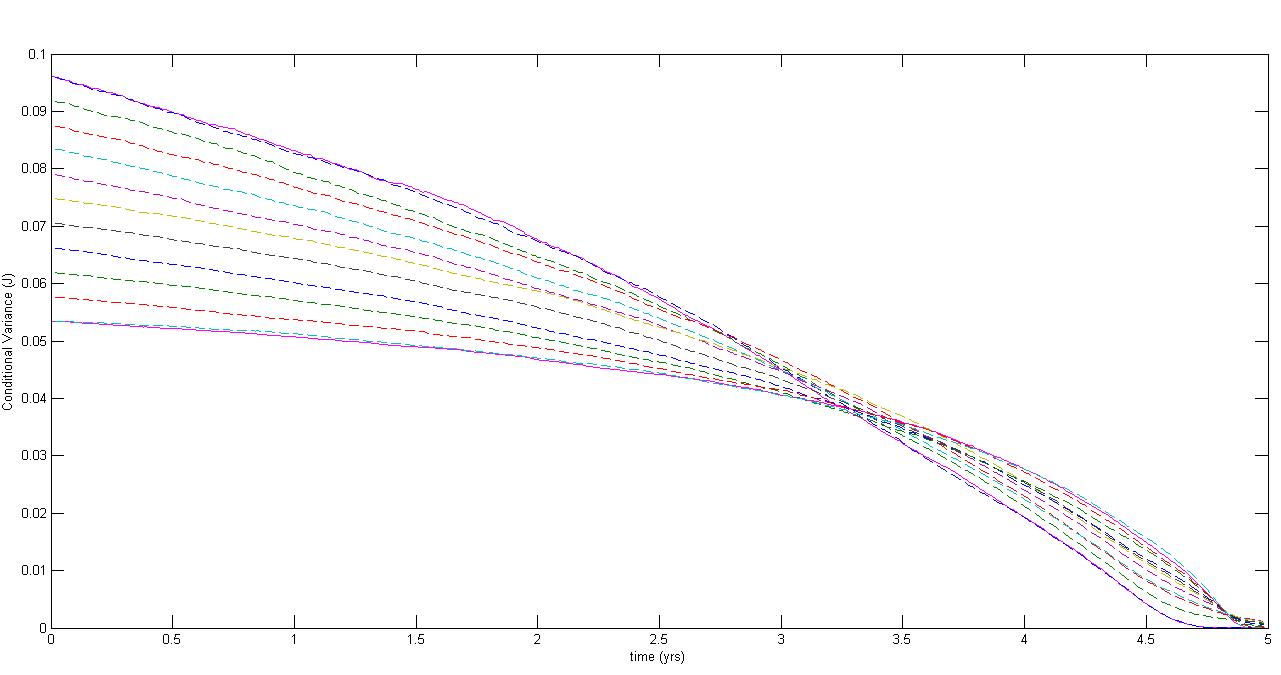}
\caption{Each line represents the average of 5,000 paths of volatility 
processes. The solid lines are that of the original BHM model with the 
top and bottom line taking $\sigma=0.9$ and $\sigma=0.5$ respectively, 
$X_T=\{0,1\}$, 
$p_i=\{0.2,0.8\}$, $r=0\%$ and $T=5$. The dashed 
lines are mean volatilities of the uncertain information model taking 
values, from top to bottom respectively, 
$\sigma_k=\{0.5,0.9\}$ with $q_k=\{0,1\}$, $q_k=\{0.1,0.9\}$, 
$q_k=\{0.2,0.8\}$,... and $q_k=\{1,0\}$.}
\label{VolProcessALL}
\end{figure}

In view of this, we investigate further the behaviour of the volatility processes. 
Because in our example here the random variable $\sigma$ is chosen such 
that it is not measurable for any $t\in[0,T]$, we can expect some degree of 
uncertainty to sustain until the very last moment. Indeed, the behaviour of the 
price process in 
the uncertain information model is not entirely counterintuitive; we expect, for 
example, that the price process behaves like a mixture of the BHM with the 
probabilities $\{q_k\}$ denoting the weights of the possible scenarios of 
$\sigma_k$. That this intuition is more or less correct, albeit there are subtle 
details, is illustrated in figure \ref{VolProcessALL}, in which the mean volatility 
${\bar\Sigma}_{tT}={\mathbb E}[\Sigma_{tT}]$ associated with various 
distributions for $\sigma$ are compared. The two solid lines represent 
${\bar\Sigma}_{tT}$ corresponding to the original BHM model, with the top 
line at $t=0$ taking the value $\sigma=0.9$ and bottom line taking the value 
$\sigma=0.5$. The dashed lines are mean volatilities generated by our 
uncertain information model. They take values $\sigma_k=\{0.5,0.9\}$ with 
the following probabilities: from the top at $t=0$, $q_k=\{0,1\}$; 
$q_k=\{0.1,0.9\}$; $q_k=\{0.2,0.8\}$; $q_k=\{0.3,0.7\}$, and so on, until 
$q_k=\{1,0\}$  for the bottom dashed line that coincides with the solid line.

The result reveals the following features: First, we see that when $\sigma=0.5$ 
the average volatility starts with a lower value as compared to that of 
$\sigma=0.9$; but eventually overrides the latter because information is 
revealed towards the end, and hence causing higher volatility. Second, we 
see that as we shift the mean of $\sigma$ from $0.9$ to $0.5$, the mean 
volatility shifts accordingly. The graph shows that the mean volatility of 
the random $\sigma$ model behaves like a weighted sum of the BHM 
mean volatilities, with weights given by the \textit{a priori} probability 
$\{q_k\}$. In this regard, the random information model studied here can 
be viewed as a BHM mixture model. 

\begin{figure}[htp]
\centering
\includegraphics[width=0.9\textwidth,clip]{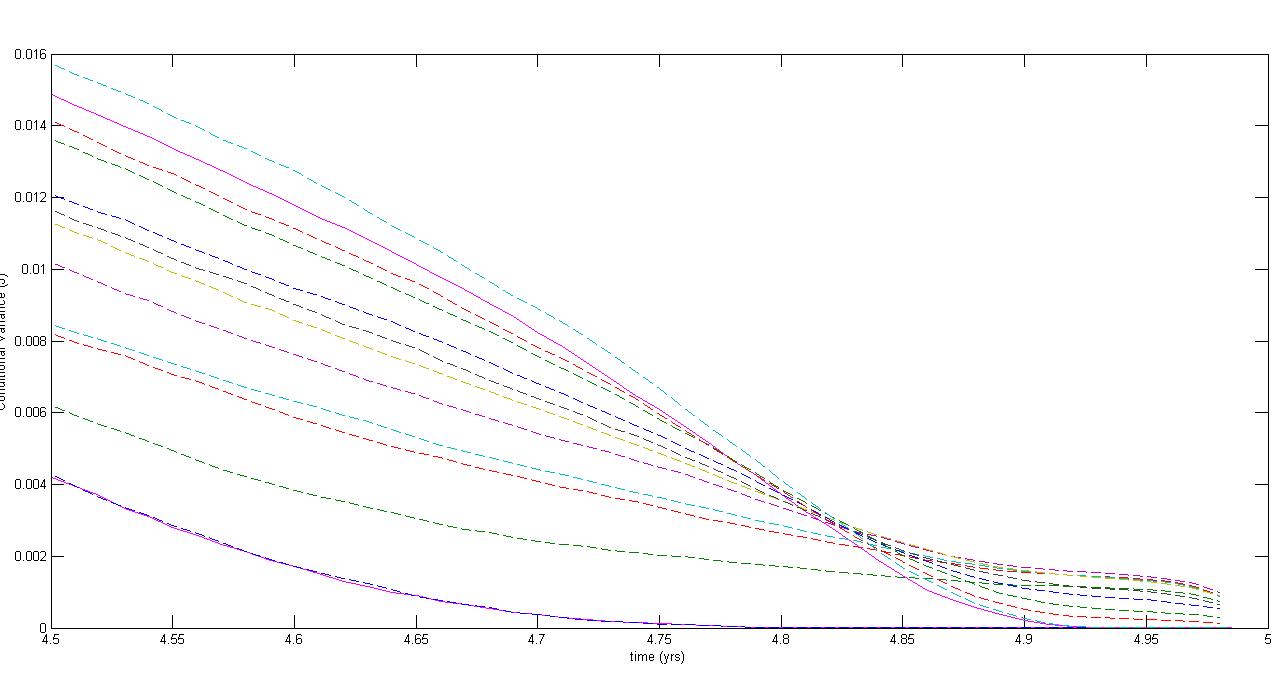}
\caption{Magnification of the averages of volatility processes. Each line 
represents the average of 5,000 paths of volatility 
processes. The solid lines are that of the original BHM model with the 
top and bottom line taking $\sigma=0.9$ and $\sigma=0.5$, respectively, 
$X_T=\{0,1\}$, $p_i=\{0.2,0.8\}$, $r=0\%$ and $T=5$. The dashed 
lines are mean volatilities of the uncertain information model taking 
values, from top to bottom respectively, 
$\sigma_k=\{0.5,0.9\}$ with $q_k=\{0,1\}$, $q_k=\{0.1,0.9\}$, 
$q_k=\{0.2,0.8\}$,... and $q_k=\{1,0\}$.}
\label{VolProcessAllZoom4550}
\end{figure}

We also draw attention to the limiting behaviour of the mean volatility as 
$t \rightarrow T$. In figure \ref{VolProcessAllZoom4550} we see clearly that 
the mean volatility of the random information model does not reach zero when 
$t$ is very close to $T$, even though that of the BHM model has. This indicates 
that there are `last minute surprises' because information regarding the terminal 
value is still uncertain, due to the additional uncertainty in $\sigma$. This is 
the essence of the random information flow rate model, as it truly describes 
the uncertainty in the market up until the last moment. Note, however, that 
values of mean volatility above $t\gtrsim 0.996T$ were not available numerically, 
due to the existence of a cancellation of large numbers that is not recognised in 
the code.

\section{Mutual information analysis}
\label{sec:7}

An alternative way of investigating properties of the price process is to 
study the behaviour of mutual information between $B_{tT}$ and $X_T$, or, 
equivalently, between $\xi_t$ and $X_T$. This is the quantity that measures 
the amount of information contained in the asset price about the value of the 
impending cash flow (cf. \cite{GY,K}), and thus represents how much the 
market has learned about the value of $X_T$ \cite{BDFH}. The mutual 
information is obtained by determining the expression:
\begin{eqnarray}
J(\xi_t , X_T) = \sum_i^n \int_{-\infty}^{\infty} \rho_{\xi X}(x, i) \ln\left( 
\frac{\rho_{\xi X}(x,i)}{\rho_{\xi} (x) \rho_X(i)} \right) \rd x, 
\end{eqnarray}
where
\begin{eqnarray}
\rho_{\xi X}(x, i) =\frac{\rd}{\rd x}\mathbb{Q} \left[ (\xi_t < x) \cap (X_T = x_i) \right]
\end{eqnarray}
is the joint density function of the random variables $(\xi_t, X_T)$, and $\rho_\xi$, 
$\rho_X$ are the respective marginal probabilities. By independence of $X_T$ 
and $\sigma$, we find that
\begin{eqnarray}
{\mathbb Q} \left[ (\xi_t < x) \cap (X_T = x_i)\right] = 
\sum_k^m \mathbb{Q}(\xi_t < x \mid X_T=x_i, 
\sigma =\sigma_k) \mathbb{Q} (X_T=x_i) 
\mathbb{Q}(\sigma= \sigma_k),
\end{eqnarray}
from which it follows that 
\begin{eqnarray}
\rho_{\xi X}(x, i) = \sum^m_k q_k p_i \frac{1}{\sqrt{2\pi t(T-t)/T}}
\exp\left( -\frac{1}{2}\frac{(x- \sigma_k x_it)^2}{t(T-t)/T} \right)
\end{eqnarray}
since conditional on $X_T=x_i$ and $\sigma=\sigma_k$, the random 
variable $\xi_t$ is normally distributed with mean $\sigma_k x_i t$ and 
variance $t(T-t)/T$. 

An alternative way of deriving the mutual information is via the formula 
\begin{eqnarray}
J(\xi_t, X_T) = H_0 - \mathbb{E}[H_t], 
\end{eqnarray}
where the Shannon-Wiener entropy $\{H_t\}$ is defined by the expression:
\begin{eqnarray}
H_t = - \sum_{i=1}^n \pi_{it} \ln\pi_{it}.
\end{eqnarray}
The entropy process $\{H_t\}_{0 \leq t < T}$ has the property that 
$\lim_{t \rightarrow T} H_t =0$. This follows from the fact that the conditional 
probability process $\{\pi_{it}\}_{0 \leq t < T}$ has the limiting behaviour
\begin{eqnarray}
\lim_{t \rightarrow T}\pi_{it}(\omega) = {\mathds 1}\{X_T(\omega) = x_i\}
\end{eqnarray}
for $i =1,\ldots,n$. To see this, suppose that for some $\omega \in \Omega$ 
we have $X_T(\omega)=x_a$ and $\sigma(\omega)=\sigma_b$ for some 
$a,b$. For this realisation the information process is given by $\xi_t=
\sigma_b t x_a+\beta_{tT}$. Substituting this expression for $\xi_t$ into the 
expression for $\pi_{it}$, and dividing the denominator and the numerator 
by the exponential factor containing $x_a$ and $\sigma_b$, we deduce 
that
\begin{eqnarray}
\pi_{at} =\frac{p_a\left( q_b+\sum_{k \neq b}q_k \exp\left[\frac{T}{T-t}
\left(x_a(\sigma_k-\sigma_b) \beta_{tT}-\frac{1}{2}x_a^2(\sigma_k-
\sigma_b)^2t\right)\right] \right)}{p_a q_b + \sum_{i \neq a}\sum_{k 
\neq b} p_i q_k \exp\left[\frac{T}{T-t}\left((x_i \sigma_k-x_a\sigma_b)
\beta_{tT}-\frac{1}{2}(x_i^2\sigma_j^2-x_a^2\sigma_b^2) t\right) \right]}.
\end{eqnarray}
As $t \rightarrow T$ all of the terms in the sums vanish. Therefore, 
$\lim_{t \rightarrow T}\pi_{at}=1$ and furthermore, since $\sum_{i} 
\pi_{it}=1$ for all $t$, we must have $\lim_{t \rightarrow T} \pi_{it}=0$ for 
$i \neq a$. Finally, since 
\begin{eqnarray}
H_t= -\ln\prod^n_{i=1} \pi_{it}^{\pi_{it}},  
\end{eqnarray}
we deduce that $\lim_{t \rightarrow T}H_t=0$. 

\begin{figure}[h!]
\centering
\includegraphics[width=0.9\textwidth,clip]{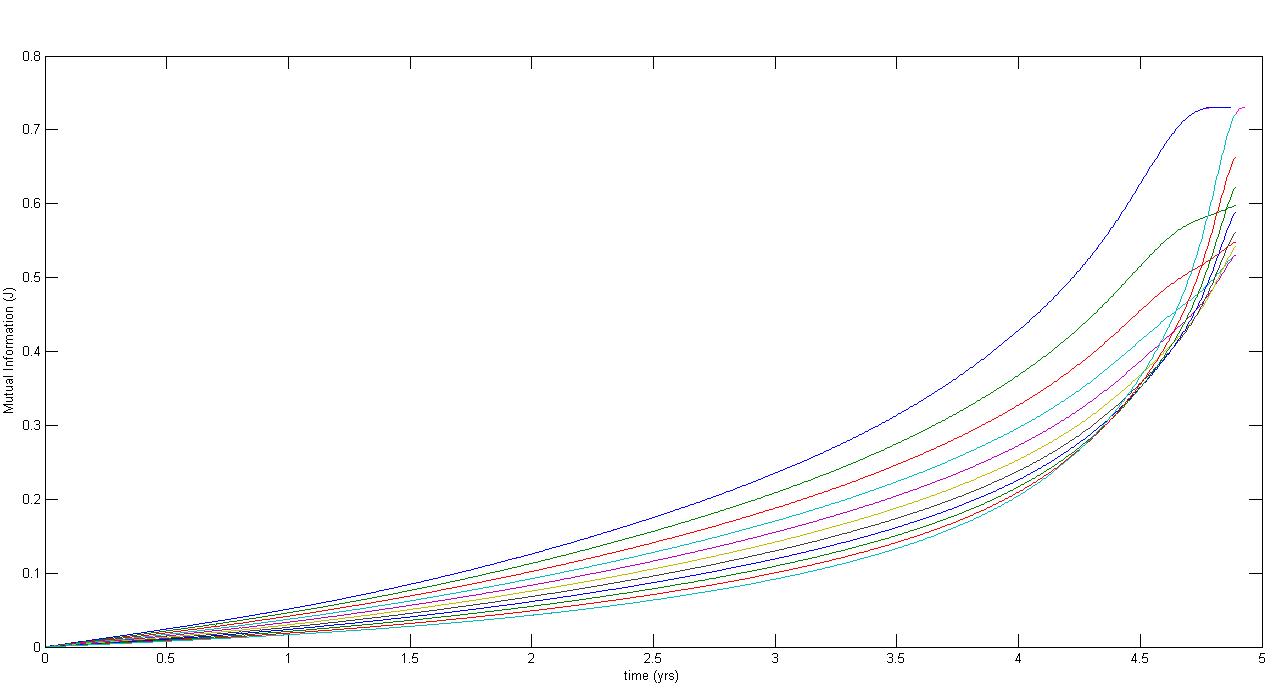}
\caption{Mutual information under the random $\sigma$ model. We let 
$\sigma_k=\{0.5,0.9\}$ with probabilities, from top to bottom at early 
times, $q_k=\{0,1\}$, $q_k=\{0.1,0.9\}$,...,$q_k=\{1,0\}$ respectively; 
$X_T=\{0,1\}$, $p_i=\{0.2,0.8\}$, and $T=5$.}
\label{MutualALLwithNaN}
\end{figure}

In figure~\ref{MutualALLwithNaN} we have a graphical illustration of the 
behaviour of mutual information. The idea here is to relate this with the 
average volatility 
that we observed in the previous section. We have used exactly the 
same parameter range to make the comparison more transparent. 
Hence, the mutual information for the BHM model with $\sigma=0.9$ 
and $\sigma=0.5$ form the upper and lower bounds, respectively, at 
early times in figure~\ref{MutualALLwithNaN}. The mutual information 
curves that lie within these bounds at early times are those under the 
uncertain information 
model, taking the values $\sigma_k=\{0.5, 0.9\}$ with probabilities 
$q_k=\{0,1\}$, $q_k=\{0.1,0.9\}$, $\cdots$, $q_k=\{1,0\}$, respectively, 
from top to bottom. 

The results shown in the figure match exactly the findings from the 
mean volatility analysis. The gap in the plots close to $t\approx T$ is 
again due to numerical limitations. 
Observe the crossover pattern seen here, as we change the distribution of 
$\sigma$. The processes with a low mutual information at early times cross 
the higher ones at later stages because the volatility is greater towards the end 
so as to `catch up' with the learning rate. Further, there are still uncertainties 
left until the very last moment. 

\section{Option on credit-risky bonds}
\label{sec:8}

We now turn to the problem of pricing options on a credit-risky bond in 
our uncertain information flow-rate extension of the BHM model. Our 
analysis follows closely that given in \cite{bhm1}. 
Consider the valuation of a European-style call option on the defaultable 
discount bond maturing at $T$. The option maturity is $t\leq T$, and the 
strike is $K$. The initial value of the call option is:
\begin{eqnarray}
C_0 &=& P_{0t} {\mathbb E}[ (B_{tT}-K)^+] \nonumber \\
&=& P_{0t} {\mathbb E}\left[ \left( \sum_i P_{tT} \pi_{it}x_i -K\right)^+ 
\right], 
\end{eqnarray}
where $\pi_{it}$ is as given in (\ref{eq:20}). If we define the positive 
process $\{\Phi_{t}\}$ according to 
\begin{eqnarray}
\Phi_t &=& \sum_i \sum_k p_{ikt}, 
\end{eqnarray}
then the call price can be expressed in the form 
\begin{eqnarray}
C_0 = P_{0t} {\mathbb E}\left[ \frac{1}{\Phi_t} 
\left( \sum_i \sum_k (P_{tT}x_i -K)p_{ikt} \right)^+ \right].
\end{eqnarray} 
Our strategy now is to eliminate the term $\Phi_t^{-1}$ via a measure 
change technique. 

We remark first that it follows from (\ref{eq:17}) and (\ref{eq:22}) that 
\begin{eqnarray}
\rd \Phi_t = \left(\frac{T}{(T-t)}\,{\mathbb E}[ \sigma X_T | \xi_t]\right)^2 
\Phi_t \rd t +\frac{T}{T-t}\, {\mathbb E}[\sigma X_T | \xi_t] \Phi_t \rd W_t, 
\end{eqnarray}
from which it follows that 
\begin{eqnarray}
\rd \Phi^{-1}_t = - \frac{T}{T-t}\, {\mathbb E}[\sigma X_T | \xi_t] 
\Phi^{-1}_t \rd W_t.  
\end{eqnarray}
Expressed in an integral form, we thus have 
\begin{eqnarray}
\Phi_t^{-1} = \exp \left( - \int^t_0 \frac{T}{T-s}\, {\mathbb E} 
[\sigma X_T | \xi_s] \rd W_s - \half \int^t_0 \frac{T^2}{(T-s)^2}\, 
{\mathbb E}[ \sigma X_T | \xi_s ]^2 \rd s \right).
\end{eqnarray}
Since ${\mathbb E}[ \sigma X_T | \xi_t]$ is bounded, the Novikov condition
\begin{eqnarray}
{\mathbb E}\left[ \exp\left(\half \int^t_0 \frac{T^2}{(T-s)^2}\, {\mathbb E}
[\sigma X_T|\xi_s]^2 \rd s \right) \right] < \infty
\end{eqnarray}
is satisfied. Hence $\{\Phi_t^{-1}\}_{0 \leq t \leq u<T}$ is a martingale. We 
also deduce that 
\begin{eqnarray}
\Phi_0^{-1} = \left( \sum_i \sum_k p_i q_k \right)^{-1}=1, 
\end{eqnarray}
and hence that ${\mathbb E}[\Phi_t^{-1}] =1$. Thus the factor $\Phi_t^{-1}$ 
can be used to effect a change of measure. Writing ${\mathbb B}_T$ for the 
new measure, the option price then becomes: 
\begin{eqnarray}
C_0 = P_{0t} {\mathbb E}^{\mathbb{B}_T} \left[ \left( 
\sum_i \sum_k (P_{tT} -K) p_{ikt} \right)^+ \right]. 
\end{eqnarray}

It is not difficult to show that under ${\mathbb B}_T$, the information process 
$\{\xi_s\}_{0 \leq s \leq t}$ is Gaussian with mean $0$ and variance $t(T-t)/T$; 
that is to say, $\{\xi_t\}$ is a ${\mathbb B}_T$-Brownian bridge. The 
Radon-Nikod\'ym derivative associated with this measure change is thus 
$\textrm{d}\mathbb{B}_T=\Phi^{-1}_t \textrm{d}\mathbb{Q}$. The random 
variables $X_T$ and $\sigma$ have the same probability law with respect to 
${\mathbb B}_T$ as with ${\mathbb Q}$, and the conditional expectation of 
any integrable function $f(X_T,\sigma)$ can be expressed as:
\begin{eqnarray}
{\mathbb E}^{\mathbb{B}_T}[f(X_T, \sigma)|{\mathcal F}_t] = 
\frac{{\mathbb E}^{\mathbb{Q}} [\Phi^{-1}_t f(X_T,\sigma)|{\mathcal F}_t]}
{{\mathbb E}^{\mathbb{Q}}[\Phi^{-1}_t | \mathcal{F}_t] }.
\end{eqnarray}
In particular, the process $\{W^*_t\}_{0 \leq t \leq u}$ defined by 
\begin{eqnarray}
W^*_t = \int^t_0 \frac{T}{T-s}\, {\mathbb E}[\sigma X_T|\xi_s]\rd s+ W_t 
\label{eq:39}
\end{eqnarray}
is a ${\mathbb B}_T$ Brownian motion. To verify that $\{\xi_t\}$ is a 
${\mathbb B}_T$-Brownian bridge,  we substitute (\ref{eq:22}) in 
(\ref{eq:39}) to deduce 
\begin{eqnarray}
\rd \xi_t = -\frac{\xi_t}{T-t}\, \rd t + \rd W^*_t.
\end{eqnarray}
But this is just the SDE for a Brownian bridge process in the 
${\mathbb B}_T$ measure. 

Returning to the problem of option pricing, let us begin by considering the 
case of a binary bond whereby the cash flow takes the two possible values 
$\{x_0,x_1\}$. Then we have 
\begin{eqnarray}
C_0 = P_{0t}\mathbb{E}^{\mathbb{B}_T} \left[ \left((P_{tT}x_1 -K) 
\sum_k p_{1kt}+ (P_{tT}x_0 -K)\sum_k p_{0kt} \right)^+ \right].
\label{OptionPrice}
\end{eqnarray}
This expectation is nontrivial when $P_{tT}x_1 > K > P_{tT}x_0$. In this case, 
the option can expire either in the money or out of the money, depending on 
whether $\xi_t>\xi^*$ or $\xi_t<\xi^*$, where $\xi^*$ is the unique critical value 
of $\xi_t$ such that $B{_tT}=K$, or, equivalently, the unique solution to the 
relation 
\begin{eqnarray}
\frac{\sum_k q_k \exp(\frac{T}{T-t}(\sigma_k x_0 \xi^* - \frac{1}{2}\sigma_k^2 
x_0^2 t))}{\sum_k q_k \exp(\frac{T}{T-t}(\sigma_k x_1 \xi^* - \frac{1}{2} 
\sigma_k^2 x_1^2 t))} = \frac{p_1 (K- P_{tT}x_1)}{p_0 (P_{tT}x_0-K)} . 
\label{eq:42}
\end{eqnarray}
Note that in general $\xi^*$ has no closed-form expression. However, the 
solution to (\ref{eq:42}) can be obtained by simple numerical root-finding 
methods. That the solution to (\ref{eq:42}) is unique (assuming that $\sigma$ 
is a positive random variable) can be seen by the fact that the bond price is 
an increasing function of $\xi_t$. 

The problem of option pricing thus reduces to performing an elementary 
Gaussian integration. We now consider the case where $X_T$ need not 
be a binary variable. The computation simplifies further if we introduce 
a standard normal variable $Z$ according to 
\begin{eqnarray} 
Z = \frac{\xi_t}{\sqrt{t(T-t)/T}} . 
\end{eqnarray}
We write $Z^*$ for the corresponding critical value. Then the option 
pricing formula is:
\begin{eqnarray}
C_0 = P_{0t}\sum_k \sum_i q_k p_i(P_{tT}x_i-K) 
N(\sqrt{\tau}\sigma_k x_i-Z^*) , 
\label{eq:44}
\end{eqnarray}
where $\tau = tT/(T-t)$ and $N(\cdot)$ denotes the standard cumulative 
normal density function.

\begin{figure}[htp]
\centering
\includegraphics[width=0.9\textwidth,clip]{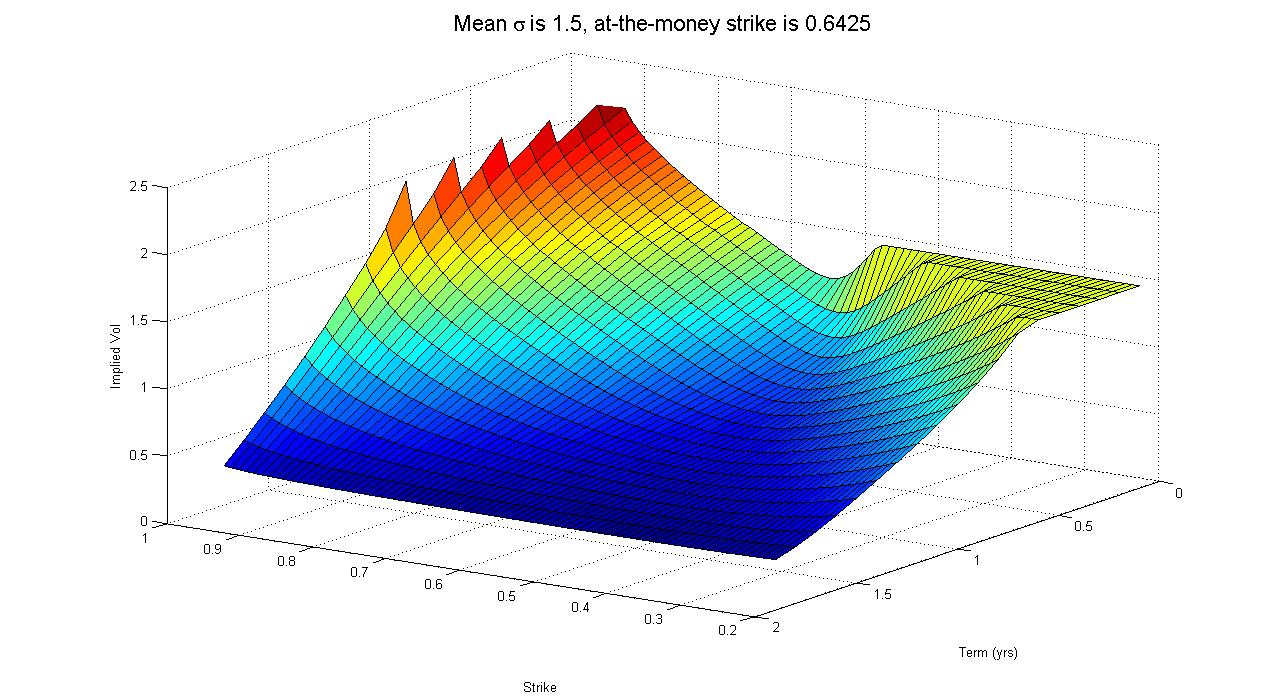}
\caption{\textit{The implied BHM volatility}. 
The implied volatility surface resulted from calibrating each strike 
and maturity of the random information model to the BHM model is 
shown. The parameters are set to be
$X_T=\{0,1\}$, $p_i=\{0.2,0.8\}$, $\sigma_{\text{BHM}}=1.5$, 
$\sigma_{\text{Rand}}= \{0.3, 2.7\}$, $q_k=\{0.5,0.5\}$, $r=0\%$ and 
$T \in [0,2]$.}
\label{VolSurfBHM}
\end{figure}

Remark: We have noted that our random information model can be 
viewed as a mixture of BHM models. In the context of the Black-Scholes 
model, Renault and Touzi \cite{RT} or Brigo \textit{et al}. \cite{BMR}, for 
instance, have carried 
out similar analysis, where the Black-Scholes volatility parameter $\sigma$ 
is taken to be time dependent random variable that is independent of the 
underlying Brownian motion. By conditioning on the volatility path, the 
European call option is obtained as the expectation of the Black-Scholes 
call price with time-averaged volatility. In particular, when $\sigma$ is 
time independent, the result of \cite{RT,BMR} for the extended 
Black-Scholes model is similar in nature to our result on extended 
BHM model. 

Remark: The randomisation of the Black-Sholes volatility parameter 
(the log-normal mixture) in the literature is carried out essentially in an 
\textit{ad hoc} manner, without any fundamental economic reason. Rather, 
it is justified on the practical ground that it gives a better handling of 
calibration. In contrast, in our model the randomisation arises from a 
more realistic analysis on the market information process. Hence, 
although the net effect is similar in both cases, our model is accompanied 
by a better justification, which in turn also gives a better justification for 
the lognormal mixture models. 

The random information-flow model considered here can be viewed as 
the simplest stochastic volatility model for option pricing in the information 
based asset pricing framework. One natural and important question arising 
in the present model is: how do we calibrate the distribution of $\sigma$? 
The answer is given by the volatility surface. We note that in the original 
BHM model the information flow rate parameter is calibrated by the option 
price for a fixed strike and a fixed maturity. Hence the model cannot be 
used to calibrate against the volatility surface. In contrast, the 
random-$\sigma$ model considered here has a wider flexibility that allows 
for the calibration of larger market data set. As an example, we have 
plotted in figure~\ref{VolSurfBHM} the option price (\ref{eq:44}) in our 
random-$\sigma$ model, but expressed in the form of an implied BHM 
volatility surface.

\section{Information manipulation}
\label{sec:9}

We conclude by drawing attention to another interesting feature of the 
variable $\sigma$ model. This concerns the notion of information 
manipulation, or, equivalently, a deliberate misrepresentation of information. 
The question 
that we are interested in here is the following: How does 
one model the manipulation of information in the information-based 
framework? One possible solution is given by the misspecification of 
the information flow-rate $\sigma$. 

The idea can be sketched as follows. Each market agent reveals information, 
expressed in the form of one of the information processes of (\ref{eq:3}). The 
impact of that agent's information on the market is then determined through 
formula (\ref{eq:4}). If an agent releases information that is based purely 
on speculation, then that information source is noise dominated, having a 
small value of the information flow rate. However, if that agent is trying to 
mislead the market, based on a reliable piece of information, then that 
information source is no longer noise dominated. Instead, this misinformation 
can be modelled by the fact that the agent provides an incorrect value for the 
information flow rate parameter. In this way, the market will estimate the 
fair price of the asset by use of the pricing formula (\ref{eq:8}), but based on 
the incorrect value for the information flow rate parameter. As a consequence, 
the market price will be misled. 

\begin{figure}[t]
\center
\includegraphics[width=0.9\textwidth,clip]{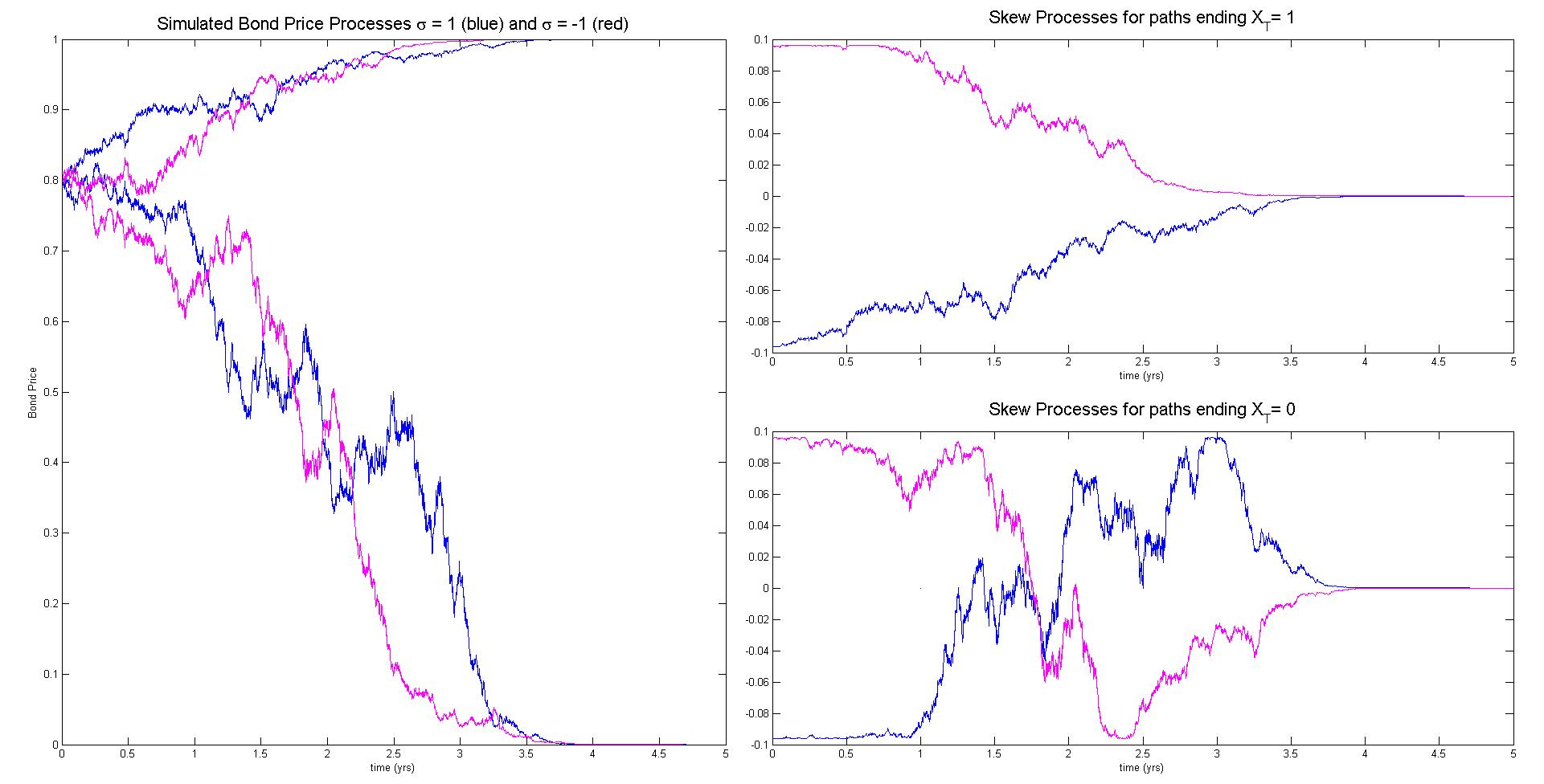}
\caption{Left: Sample paths of the BHM with $\sigma=1$ (blue) and 
$\sigma =-1$ (red) using same Brownian Bridges; and the corresponding 
skew processes associated with $X_T(\omega)=1$ (Top right) and 
$X_T(\omega)=1$ (Bottom right). Here we set $X_T=\{0,1\}$, 
$p_i=\{0.2,0.8\}$, $r=0\%$ and $T=5$.}
\label{NegativeSigma}
\end{figure}

As an illustration of this behaviour, in figure~\ref{NegativeSigma} we show 
sample paths for the defaultable digital bond price process; in one case where 
the bond did not default, whereas in the other case where the bond defaulted. 
In each case, two sample paths are given; one corresponding to where the true 
market price ought to be if there is no misleading information, and one 
corresponding to where the realised price is, owing to the existence of a 
deliberate price manipulation. In order to make the effect of price manipulation 
visually pronounced, here we have taken a slightly extreme case in which the 
true value of the information flow rate is $\sigma =+1$ (blue), whereas the 
`conjugate' price process is generated by the false belief that the flow rate is 
given by $\sigma=-1$ (red). We find in these examples the existence of a kind 
of anti-correlation between the `true' and `false' prices around their conditional 
means. This is illustrate more clearly in the skewness plot, also shown in 
figure~\ref{NegativeSigma}. 

One might enquire in which way a mis-specification of the information 
flow rate $\sigma$ is realised in practice. In this connection it is worth 
remarking that there is an extended literature on price manipulation, often 
in the context of insider trading. One typical way of spreading a false 
information is by taking a trading position in a strategic manner (cf. 
\cite{ohara,AG}). For example, suppose that an informed trader has the 
information that the price of an asset is likely to drop in near future. In this 
case, taking a short position amounts to effectively revealing the content of 
that information. Hence, by momentarily taking a long position before 
taking a short position, an informed trader can mislead the market. One 
can think of such a deliberate manoeuvre being represented abstractly 
in the form of one of the information processes in (\ref{eq:3}) taking an 
`incorrect' value of $\sigma$. 

We see therefore that the information-based framework allows for a range 
of flexible extensions to model various scenarios that might occur in a 
given financial market. It would be of considerable interest, in particular, 
to develop further the information-based approach to price manipulation.

\begin{acknowledgments}
The authors thank M.~H.~A.~Davis and L.~P.~Hughston 
for comments and stimulating discussion. YTL thanks HSBC for 
support. 
\end{acknowledgments}

\end{document}